# Probing Phase Fluctuations in Cuprates by Multidimensional Josephson Spectroscopy

A. Liu[1,2,*], D. Pavicevic[2], A. G. Salvador[3], M. H. Michael[2], M. Fechner[2],
P. M. Lozano[1,4,†], Q. Li[1,4], G. D. Gu[1], E. Demler[3], A. Cavalleri[2,5,*]

[1]*Condensed Matter Physics and Materials Science Division, Brookhaven National Laboratory, New York, USA*
[2]*Max Planck Institute for the Structure and Dynamics of Matter, Hamburg, Germany*
[3]*Institute for Theoretical Physics, ETH Zurich, Zurich, Switzerland*
[4]*Department of Physics and Astronomy, Stony Brook University, New York, USA*
[5]*Department of Physics, University of Oxford, Oxford, United Kingdom*
[†]*Current Address: Advanced Photon Source, Argonne National Laboratory, Argonne, IL 60439, USA*

[*]Corresponding authors: aliu1@bnl.gov; andrea.cavalleri@mpsd.mpg.de

**The normal state of cuprate superconductors is believed to host precursor electron pairing above the superconducting transition temperature. These pre-formed pairs, which extend over an unknown temperature range, are expected to lack long-range phase coherence. Because spectroscopic evidence for this hypothesis is only indirect, new methods to probe these physics are needed. Nonlinear terahertz spectroscopy allows one to probe high-order time correlation functions and is therefore sensitive to fluctuations. Measurements of third-harmonic generation along the superconducting planes have revealed responses persisting above $T_c$, but a clear relation between these optical responses and specific properties of the fluctuating state has not yet been established. Here, we apply angle-resolved multidimensional terahertz spectroscopy to probe the optical nonlinearity of the Josephson tunneling resonance in $La_{1.83}Sr_{0.17}CuO_4$ across the superconducting to normal state transition. We find that among all third-order nonlinear responses, the non-rephasing multidimensional signal is especially sensitive to non-mean field fluctuations of the order parameter. We measure a crossover occurring at $T \sim T_c$ between a mean-field optical nonlinearity that dominates at low temperatures and one that is mediated by short-range superconducting fluctuations. We present a theoretical framework that connects the out-of-plane optical response in both regimes to the fluctuating in-plane superconducting state, and contrast the sensitivity of different types of terahertz nonlinearities to fluctuating order.**

The nature of the normal state in cuprate superconductors has long presented a series of elusive challenges for theory and experiment. A recurring theme posits the presence of precursor electron pairing [1, 2] without the necessary long-range phase coherence to sustain macroscopic condensation [3]. Over the years, evidence of such pre-formed Cooper pairs and of a transition determined by an order-disorder transition of the order-parameter phase has accumulated. Early on, Uemura and coworkers noted that superfluid density scales (linearly) with $T_c$ on the underdoped side of the phase diagram [4], a simple behavior reminiscent of short-range Bose-Einstein Condensation (BEC) rather than that predicted by BCS theory, for which one expects $T_c$ to scale pairing strength and gap size. Other experimental probes such as Nernst effect [5], torque magnetometry [6], and terahertz spectroscopy [7, 8] have also yielded circumstantial evidence of local pairing. Shot noise measurements in tunnel junctions that exceed expectations for single charge tunneling [9] are also consistent with this view. Yet a spectroscopic observable that directly probes these putative short-range superconducting fluctuations [10] has so far been lacking.

A potential window into the nature of the superconducting transition arises from interlayer tunneling modes between the superconducting copper-oxygen planes. Because these planes are also capacitively coupled, the interlayer (inductive) tunnelling becomes manifest in a finite frequency Josephson plasma resonance (JPR) [11, 12]. As the tunnelling energy is proportional to $I_c$, the JPR frequency $\omega_p$ scales with superfluid density $N_s$ ($\omega_p^2 \propto N_s$) [13], and has been used as a direct indication for the onset of superconducting order in cuprates [14, 15].

The JPR frequency is typically probed by linear optical spectroscopy, in which a driving field $E_{in}$, resonant with $\omega_p$, excites Josephson plasmons at the bottom of their energy dispersion that radiate a linear optical response $\chi^{(1)}E_{in}$ (illustrated in Fig. 1a). This process dominates the interaction of light and the JPR at temperatures well below $T_c$, at which a mean-field description of the superconducting phase holds. As such, the linear optical response should likewise vanish along with the superfluid density at $T_c$, at which the bottom of the JPR energy dispersion at zero momentum is suppressed to zero.

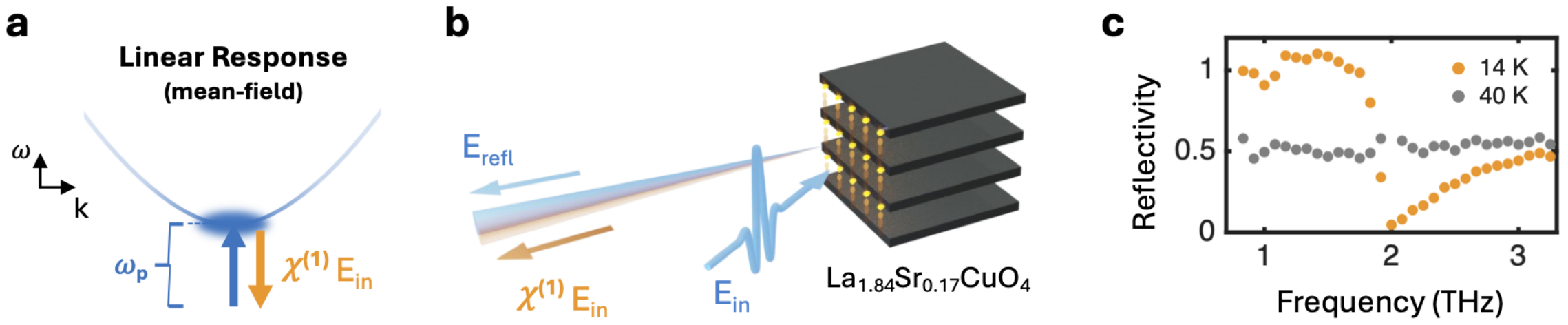

**Fig. 1. Linear optical response of the Josephson plasma resonance. a.** Schematic linear excitation of the Josephson plasma resonance by an excitation field $E_{in}$ resonant to the Josephson plasma frequency $\omega_p$, resulting in a radiated linear optical response $\chi^{(1)}E_{in}$. **b.** Experimental schematic of a linear reflectivity measurement, in which a terahertz pulse $E_{in}$ excites the Josephson plasma resonance, resulting in a linear optical response $\chi^{(1)}E_{in}$ emitted in the same direction as the reflected excitation field $E_{refl}$ that interfere to produce the observed reflectivity spectrum. **c.** Reflectivity spectrum measured at two different temperatures T = 14 K ($0.38T_c$) and T = 40 K ($1.1T_c$), obtained from time-resolved terahertz spectroscopy by scaling the ratio between the reflectivity at 45 K by the measurements in [13]. The reflectivity exhibits a well-defined edge in the superconducting phase, and is featureless above $T_c$.

The linear JPR response of optimally-doped $La_{1.83}Sr_{0.17}CuO_4$ ($LSCO_{0.17}$), with a transition temperature $T_c \approx 36.5$ K, was probed here for reference using time-domain terahertz spectroscopy. An experimental schematic for such a measurement is shown in Fig. 1**b**, in which a weak incident field $E_{in}$ (polarized perpendicular to the superconducting planes) impinges on the sample surface and induces a time-dependent phase-difference $\varphi$ between adjacent superconducting order parameters. The dynamics of $\varphi$ are that of a coordinate in a sinusoidal `washboard' potential $U(\varphi) \propto -\cos(\varphi)$, governed by the sine-Gordon equation [11], and result in a radiated field described by an effective linear susceptibility $\chi^{(1)}$. For the purposes of this discussion it is useful to consider the nature of JPR reflectivity spectra, which are fundamentally determined by an interference between the reflected field $E_{refl}$ and the field re-radiated by the linearly excited JPR polarization $P=\chi^{(1)}E_{in}$, which in turn carries a phase shift determined by the complex linear susceptibility $\chi^{(1)}=1+\varepsilon_1(\omega)+i\varepsilon_2(\omega)$. Measurements of the reflectivity at temperatures below and above $T_c$ are shown in Fig. 1**c**. Deep into the superconducting phase (T = 14 K), a pronounced reflectivity edge is observed at 2 THz [13]. Directly above the transition temperature (T = 40 K) this edge disappears and the reflectivity becomes featureless in this frequency range. In the case of strongly fluctuating superconductivity however, interlayer Josephson coupling should in principle persist even through the loss of long-range in-plane phase coherence.

In contrast to linear spectroscopy, third-order nonlinear optical spectroscopy probes higher-order time correlations, which reflect interactions and fluctuations. In-plane Third-Harmonic Generation (THG) [8] and Multidimensional Terahertz Spectroscopy (MDTS) [16] have revealed optical nonlinearities persisting above $T_c$, but an unambiguous connection of these responses to specific properties of the fluctuating superconducting state has proven difficult, due in part to the in-plane optical response being dominated by electronic quasiparticles and containing multiple competing nonlinear contributions.

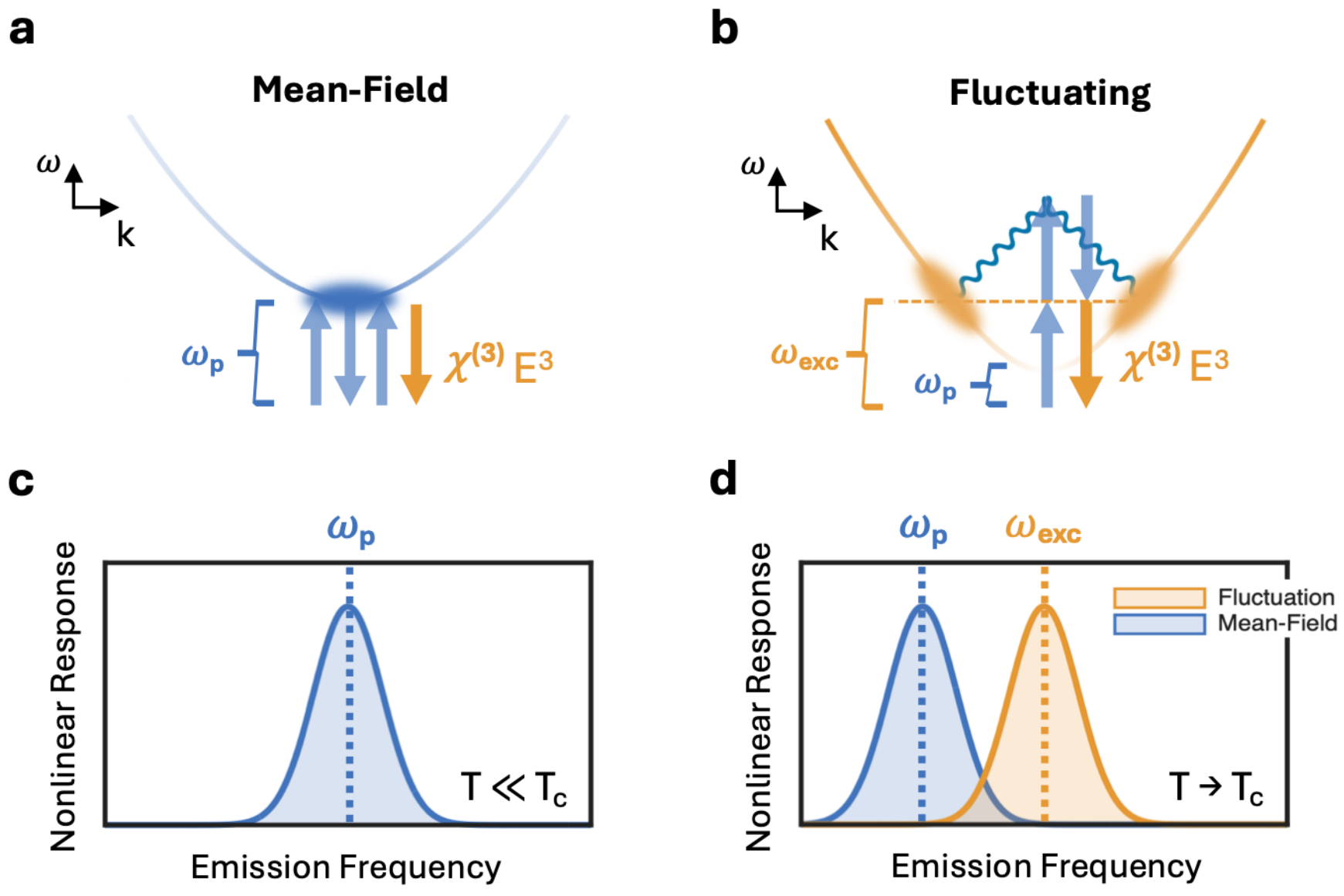

**Fig. 2. Nonlinear optical response of the JPR. a.** Schematic nonlinear excitation of the JPR by three excitation fields $E^3$ resonant to the JPR frequency $\omega_p$, resulting in a radiated nonlinear optical response $\chi^{(3)}E^3$. **b.** Additional fluctuation-mediated nonlinear optical response involving parametric excitation of finite-momentum JPR modes with two excitation fields, at energies determined by the excitation frequency $\omega_{exc}$. A third excitation field stimulates the emission of a radiated nonlinear optical response $\chi^{(3)}E^3$, again at a frequency $\omega_{exc}$. **c.** Schematic emission spectrum illustrating that at temperatures well below the transition temperature ($T \ll T_c$), a predominately mean-field nonlinear optical response results in nonlinear signal emission at the JPR frequency $\omega_p$. **d.** Schematic emission spectrum for temperatures near the phase transition ($T \to T_c$), at which thermal occupation of finite-momentum JPR modes lead to increasing fluctuation-mediated nonlinearities. These fluctuation-mediated nonlinearities can be distinguished from their mean-field counterparts by nonlinear signal emission at the excitation frequency $\omega_{exc}$.

On the contrary, nonlinear optical excitation along the out-of-plane c-axis is expected to distill signatures of the superconducting phase through tunnelling fluctuations. At lowest order, third-order optical nonlinearities measured along the c-axis derive from three-field excitations that jointly excite the JPR [17, 18, 19, 20]. The mean-field third-order optical response (analogous to the linear optical response process in Fig. 1**a**) is illustrated in Fig. 2**a**, in which the three fields excite the JPR at near-zero momentum and at its resonance frequency $\omega_p$. In such a mean-field description the third-order nonlinear polarization $P = \chi^{(3)}(\omega_1, \omega_2, \omega_3)E^3$ is comprised of both signals at the third-harmonic frequency ($\omega_1 + \omega_2 + \omega_3 = 3\omega_{exc}$) and fundamental frequency ($\pm\omega_1 \pm \omega_2 \pm \omega_3 = \omega_{exc}$). While the former signals emitting at $3\omega_{exc}$ are well-known from third-harmonic generation, the more general 'four-wave mixing' signals emitting at $\omega_{exc}$ each offer their own unique capabilities (an example of which is the rephasing (Josephson Echo) signal that was recently used to probe superconducting disorder [21, 22, 23]). In general, because the frequency of the JPR softens to zero when crossing $T_c$, this channel is expected to soften with temperature and is insensitive to short range phase fluctuations.

One additional parametric excitation process has recently been predicted [17] to emerge in the strongly fluctuating regime. This process is shown in Fig. 2**b**, in which two fields excite pairs of Josephson plasmons at finite and opposite momenta determined by the driving frequency, after which a third field $E_3$ stimulates the emission of a third-order fluctuation-mediated nonlinearity $\chi^{(3)}E_1E_2E_3$.

Importantly, mean-field and fluctuation-mediated nonlinearities result in distinct spectral signatures in experiments [17]. A mean-field nonlinear optical response results in signal emission at the JPR frequency $\omega_p$, which is the dominant process at temperatures well below $T_c$ as shown in Fig. 2**c**. In contrast, a fluctuation-mediated nonlinear optical response will result in signal emission at a frequency determined by the excitation frequency $\omega_{exc}$, as required by energy-momentum conservation in Fig. 2**b**. As temperature approaches $T_c$ and thermal occupation of the finite-momentum JPR modes increases, a crossover between dominant mean-field and fluctuation-mediated nonlinearities is expected to occur, as illustrated by the two co-existing spectral peaks centered at $\omega_p$ (mean-field) and $\omega_{exc}$ (fluctuation-mediated) in Fig. 2**d**.

Similar to what was discussed in reference [21], Multidimensional Terahertz Spectroscopy (MDTS) [24, 25, 26] was used to isolate and distill the constituent components of the nonlinear JPR response [27]. The experiment is shown schematically in Fig. 3**a**, in which two terahertz frequency pulses of light denoted $E_A$ and $E_B$ excite $LSCO_{0.17}$ in a non-collinear reflection geometry and cooperatively generate an emitted nonlinear electric field $E_{NL} = \chi^{(3)}(\omega_1, \omega_2, \omega_3)E^3$. We note that the center of our excitation spectrum ($\omega_{exc}$ = 2 THz, shown inset in Fig. 3**a**) is chosen to coincide with the low-temperature JPR frequency. By measuring $E_{NL}$ as a function of the inter-pulse delay $\tau$ and the signal emission time t, subsequent Fourier transform along the two time axes $\{\tau, t\}$ returns a two-dimensional spectrum that correlates the frequencies of coherent dynamics initiated by each pulse. Due to the difference in in-plane momenta of $E_A$ and $E_B$, different nonlinearities are emitted with unique wavevectors determined by momentum conservation [21, 28], allowing for the measurement of individual nonlinear responses by their unique phase-matched emission wavevector.

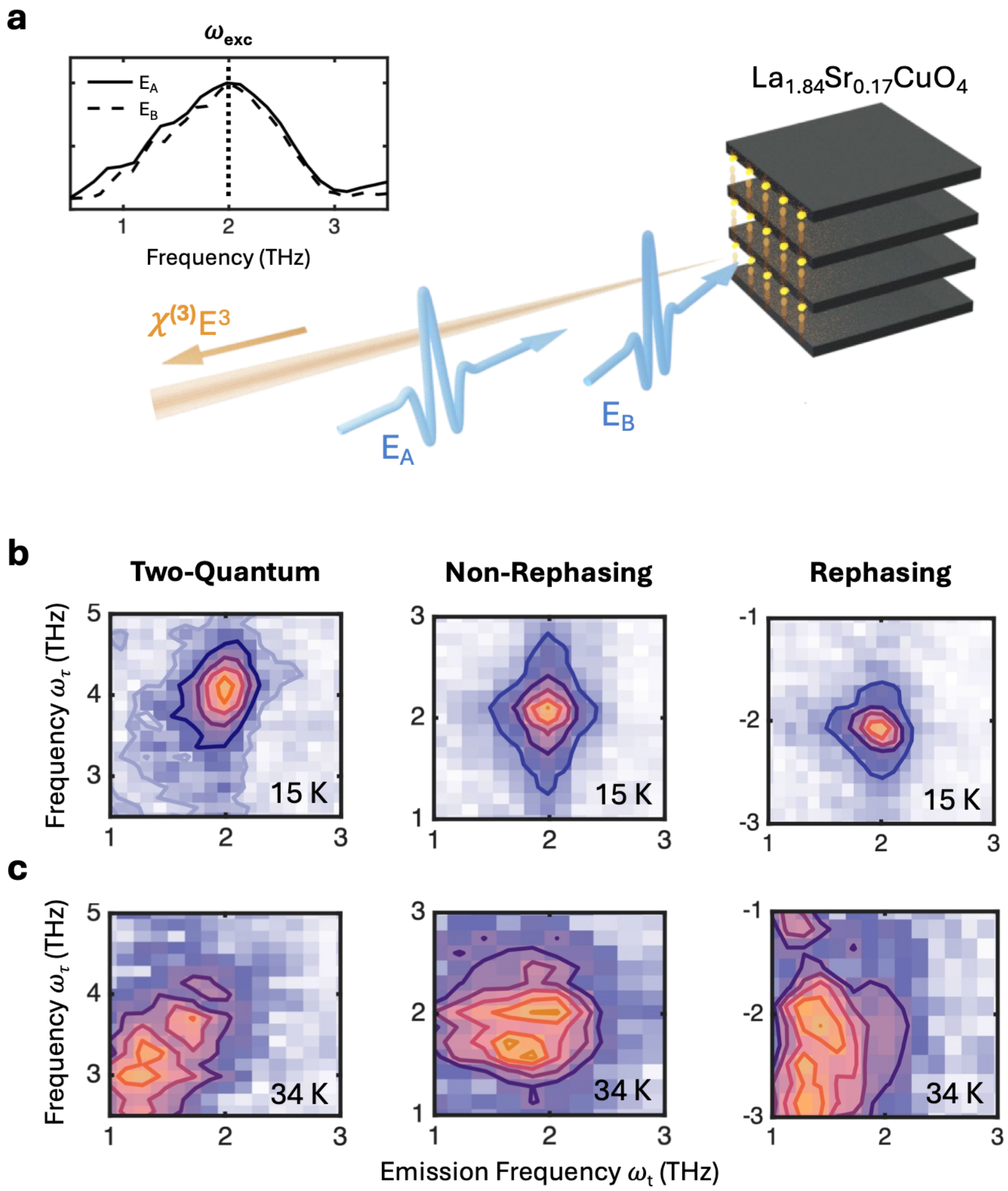

**Fig. 3. Multidimensional terahertz spectroscopy of the JPR approaching $T_c$. a.** Experimental schematic of the angle-resolved MDTS experiment (same as described in [21]), in which two excitation fields denoted $E_A$ and $E_B$ excite $LSCO_{0.17}$ along its c-axis, resulting in emitted nonlinear optical signals $\chi^{(3)}E^3$ emitted with unique wavevectors determined by wavevector phase-matching. The excitation spectra $E_A$ and $E_B$ are shown inset, with a center frequency $\omega_{exc}$ = 2 THz. **b.** MDTS spectra of the three coherent four-wave mixing signals measured at low temperature (T = 15 K), exhibiting sharp peaks centered at the JPR frequency $\omega_p$. **c.** Corresponding MDTS spectra measured near the phase transition (T = 34 K), which exhibit broader and more complicated peak lineshapes. Notably, the two-quantum and rephasing signals shift in frequency space while the non-rephasing signal remains largely stationary relative to its low-temperature position.

At low temperatures, superconductivity is robust and the nonlinear JPR response should be of mean-field character, in which the third-order optical susceptibility descends directly from the linear optical response [17]. In Fig. 3**b** we show MDTS measurements at a temperature of $T = 0.41T_c$ (15 K) in accordance with this expectation. Three coherent four-wave mixing signal channels are possible, termed two-quantum, non-rephasing, and rephasing [25, 17, 18]. Here, we neglect the incoherent pump-probe signal which is increasingly influenced by coupling to the electronic quasiparticle background as temperature approaches the phase transition [25]. All three coherent signal channels exhibit peaks with positions determined by the JPR frequency $\omega_p$ along both axes and linewidths comparable to that of the JPR loss function at this temperature.

In Fig. 3**c** we show MDTS measurements at a temperature $T = 0.93T_c$ (34 K) close to the phase transition, which reveal a marked difference between the three coherent four-wave mixing nonlinearities. While the peak positions of the two-quantum and rephasing signals shift with the softened JPR frequency measured in linear spectroscopy, the non-rephasing peak remains largely stationary and broadens symmetrically. In fact, the non-rephasing peak lineshape begins to follow that of the excitation spectrum (shown inset in Fig. 3**a**) along both frequency axes, rather than reflecting signatures of the JPR loss function, indicative of a non-resonant response in this regime [29].

To clarify these peak behaviors, we integrate each peak along the vertical frequency axis $\omega_\tau$ to resolve the spectral content of their nonlinear signal emission. The integrated peak lineshapes are shown in Fig. 4, which emphasizes the drastic difference between each temperature regime. At T = 15 K all three signals emit at the low-temperature JPR frequency $\omega_t = 2$ THz, consistent with a dominant mean-field optical nonlinearity. Near $T_c$ (T = 34 K), however, a crossover into fluctuation-mediated nonlinearities begins to occur and the integrated emission lineshapes take on a bimodal character, as emphasized by the two underlying shaded peaks in Fig. 4 centered at $\omega_t = 1.35$ THz (mean-field) and $\omega_t = 2$ THz (fluctuation-mediated). We see clearly that the spectral weight shifts to lower frequency for the two-quantum and rephasing signals, while spectral weight is retained at the excitation frequency for the non-rephasing signal. This indicates that the non-rephasing signal is most sensitive to fluctuation-mediated optical nonlinearities of the type illustrated in Fig. 2**b**, while the two-quantum and rephasing signals retain a predominately mean-field character (Fig. 2**a**).

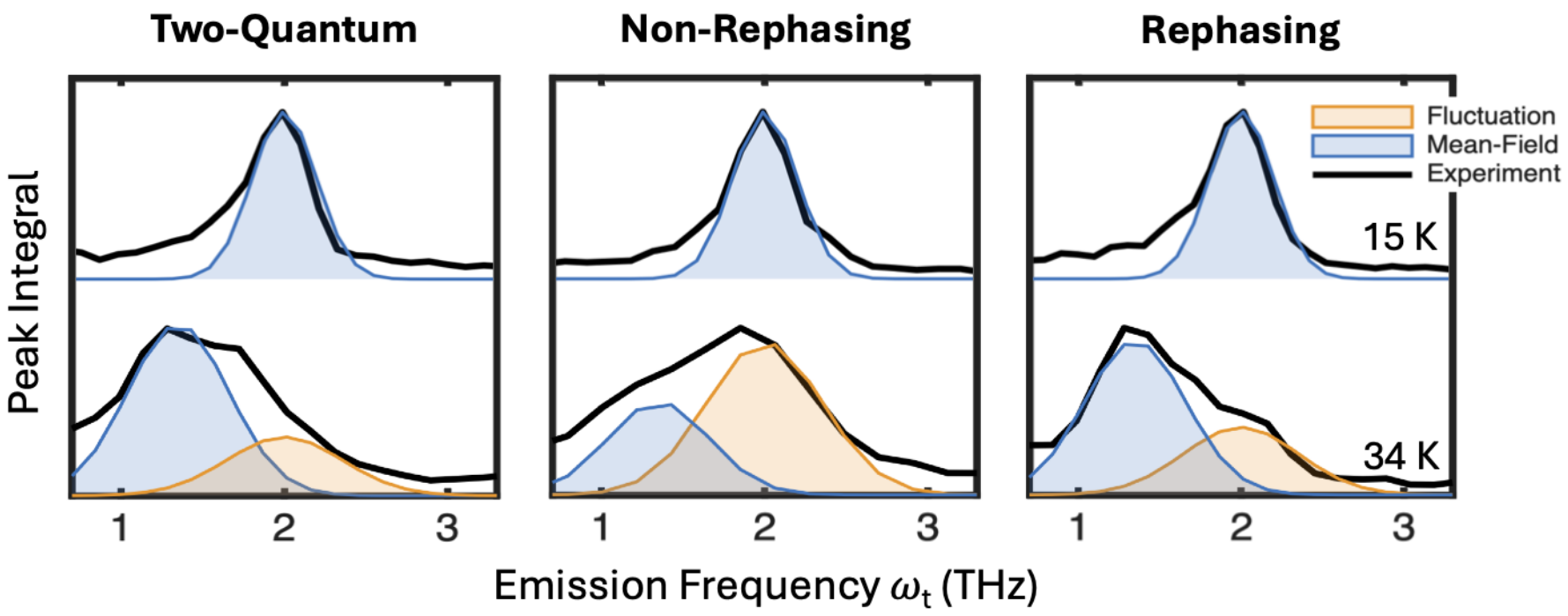

**Fig. 4. Integrated peak lineshapes reveal a crossover between mean-field and fluctuation nonlinearities.** Nonlinear emission peak lineshapes obtained by integrating the peak in each MDTS spectrum shown in Fig. 3 along the vertical frequency axis $\omega_\tau$. At a temperature of 15 K (top row), the emission lineshape is predominately mean-field in character as indicated by the single peak shaded in blue. At a temperature of 34 K (bottom row) the emission lineshape becomes bimodal, indicating the additional contribution of fluctuation-mediated optical nonlinearities. The two-quantum and rephasing lineshapes exhibit stronger mean-field contributions (blue peak, $\omega_t = 1.35$ THz) in comparison to the fluctuation-mediated contributions (orange peak, $\omega_t = 2$ THz), while the non-rephasing lineshape exhibits a stronger fluctuation-mediated contribution instead.

If the non-rephasing signal is indeed sensitive to superconducting fluctuations, we could expect it to persist even through the loss of superconducting order. We first examine this possibility theoretically by describing these physics using a fluctuating Lawrence-Doniach theory (see Supplemental Information) that connects the nonlinear c-axis optical response to fluctuations of the superconducting order parameter within the $CuO_2$ planes and to their Josephson coupling between layers. The Lawrence-Doniach free energy is

$$F_{LD} = \sum_n \int d^2r \left[a(T)|\Psi_n(\boldsymbol{r},t)|^2 + \frac{b}{2}|\Psi_n(\boldsymbol{r},t)|^4 + K_{ab}|\nabla\Psi_n(\boldsymbol{r},t)|^2 + J_\perp|\Psi_{n+1}(\boldsymbol{r},t) - \Psi_n(\boldsymbol{r},t)|^2\right],$$

where $a(T)$ is the quadratic Ginzburg-Landau coefficient, $b > 0$ stabilizes the ordered phase, $K_{ab}$ is the in-plane stiffness, and $J_\perp$ is the interlayer Josephson stiffness. The first three terms describe the Ginzburg-Landau free energy of each layer, while the last term favors phase and amplitude locking between neighboring layers. Above $T_c$, superconducting correlations are described by fluctuations of the order parameter around zero, which we treat within a Gaussian time-dependent Ginzburg-Landau theory. In this framework (and crucially taking into account the finite pulse duration of our excitation pulses; see Supplemental Information) we predict a fluctuation-mediated nonlinear optical response above $T_c$, whose constituent signals are plotted in Fig. 5**a** as a function of distance from the transition temperature $\delta = a(T)/2J_\perp$. We find that the four-wave mixing (non-rephasing, rephasing, and two-quantum) signals all exhibit an enhanced response when approaching $T_c$ from above, with strengths that follow the ratio 4:2:1 (resembling the signals observed below $T_c$ in Fig. 4). Notably, the third-harmonic signal lacks the same sensitivity to criticality, as we explain in the following.

The sensitivity of the four-wave mixing nonlinear signals to superconducting fluctuations can be understood from the frequency structure of the underlying nonlinear optical susceptibility. Consider, for example, the non-rephasing signal with frequency insertions $(\omega_1, \omega_2, \omega_3) = (\omega, -\omega, \omega)$ whose oppositely-signed frequencies $\omega_2$ and $\omega_3$ produce an effectively static perturbation of the zero-frequency superconducting fluctuations. This can be intuitively understood from its corresponding c-axis optical conductivity,

$$\sigma^{(3)}_{c,NR}(\omega,-\omega,\omega) \propto \frac{1}{\omega^2}\frac{\partial\sigma^{(1)}_c(\omega,\delta)}{\partial\delta} \propto \frac{1}{\omega^2\sqrt{\delta}},$$

which shows how the finite-frequency conductivity changes when the soft fluctuation mass is shifted. Clearly, this mechanism is not possible for the third-harmonic generation signal with the frequency insertions $(\omega_1, \omega_2, \omega_3) = (\omega, \omega, \omega)$, consistent with previous c-axis THG measurements that did not observe an above-$T_c$ response in $LSCO_{0.15}$ [30].

In Fig. 5**b**, we show MDTS spectra of the non-rephasing signal evolving across the transition temperature that shows the non-rephasing signal persisting not just to the transition temperature $T_c$, but even above the superconducting transition. At both temperatures $T \approx T_c$ (36.5 K) and $T \approx 1.1T_c$ (40 K), the spectra exhibit a peak that, again, is now entirely centered at the excitation frequency $\omega_{exc}$ as expected from the parametric nonlinearity illustrated in Fig. 2**b** and the simulated non-rephasing spectrum shown inset in Fig. 5**a**. At these same temperatures, the two-quantum and rephasing signals become vanishingly weak as the JPR oscillator strength decreases rapidly due to softening of the JPR frequency, while also shifting the JPR frequency outside our excitation bandwidth.

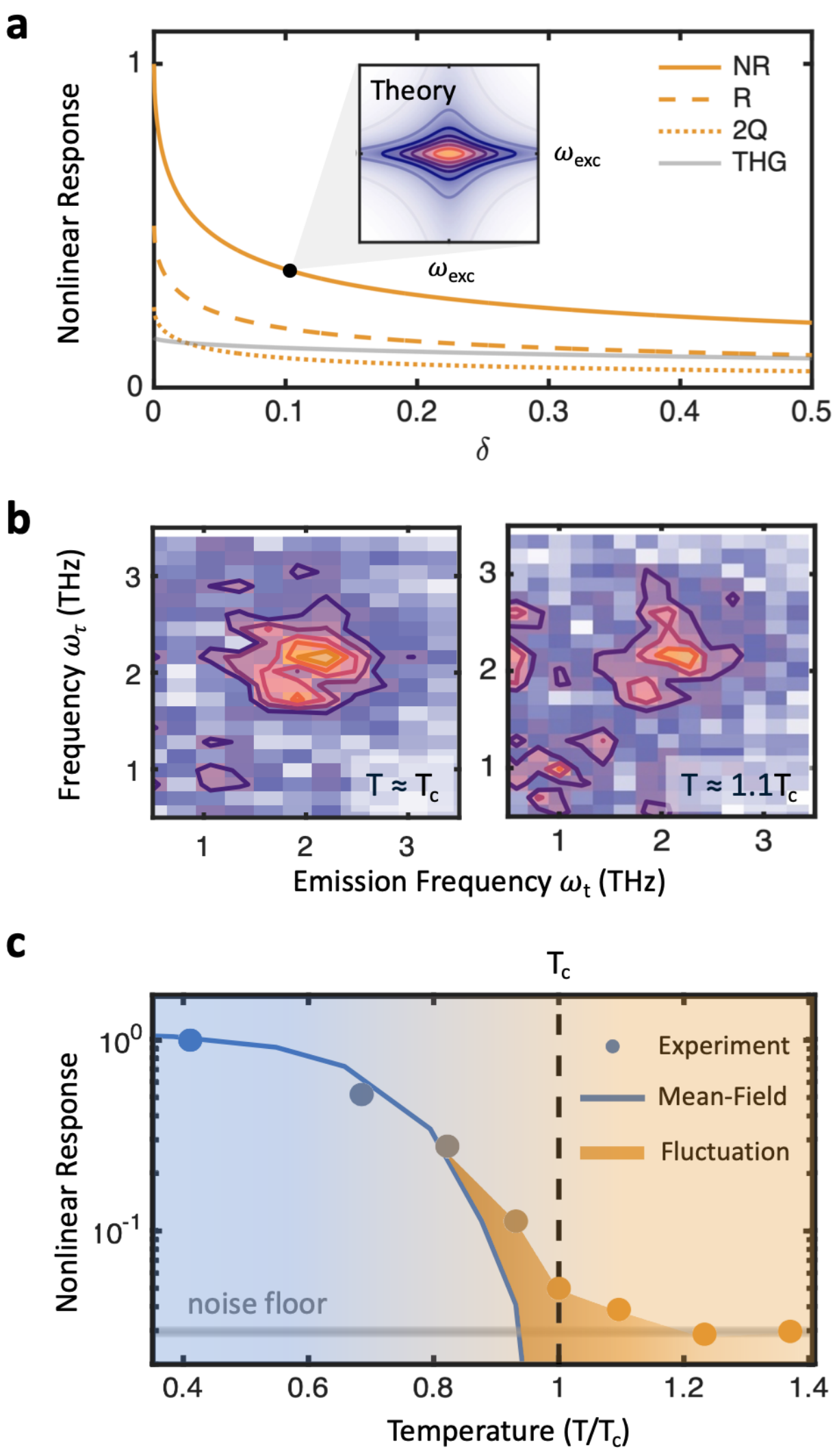

**Fig. 5. Fluctuation-mediated optical response in the normal state. a.** Theoretical prediction of fluctuation-mediated coherent four-wave mixing signals in the normal state ($T > T_c$). A critical enhancement of non-rephasing (NR), rephasing (R), and two-quantum (2Q) signals is observed, with signal strengths following the ratio 4:2:1 respectively. A simulated non-rephasing spectrum for $\delta = 0.1$ is shown inset, comprised of a single peak centered at the excitation frequency $\omega_{exc}$ along both frequency axes. Notably, the third-harmonic generation (THG) signal exhibits no critical enhancement. **b.** MDTS spectra of the non-rephasing signal at temperatures of $T = T_c$ (36.5 K) and $T = 1.1T_c$ (40 K), both exhibiting a single peak centered at the excitation frequency that agrees with the simulated spectrum in **a**. **c.** Temperature dependence of the integrated non-rephasing peak area, which exhibits a rapid decay upon approaching $T_c$ due to softening of the JPR frequency (oscillator strength) and a tail extending above $T_c$ due to the fluctuation-mediated optical response.

Finally, we examine the temperature dependence of the peak area shown in Fig. 5**c**, which exhibits the sharp decrease expected from the depletion of superfluid density but also a tail extending above $T_c$ that is characteristic of fluctuating quantum systems. These data are plotted alongside the expected (normalized) peak area in a mean-field approximation, found from an assumed $1 - (T/T_c)^4$ dependence [31] of the superfluid density and the excitation field strength at $\omega_p$. We observe a clear crossover in which the experimental data begin deviating from the mean-field expectation at around $T = 0.8T_c$, until the signal decays to the noise floor by $T = 1.2T_c$. One way to intuit this behavior is in terms of a crossover between 3-D fluctuations that extend coherently across several layers and decoupled 2-D fluctuations above and below a crossover temperature $T_{cross}$. For optimally-doped LSCO our theory estimates a crossover temperature $T_{cross} - T_c \simeq 0.54$-$1.44$ K, which provides a natural explanation for the disappearance of the signal in Fig. 5**c** in terms of the spatial dimensionality of superconducting fluctuations.

In summary, we have applied multidimensional terahertz spectroscopy to the superconducting transition in a cuprate superconductor and demonstrate Josephson four-wave mixing as a sensitive probe of superconducting fluctuations. These experiments pave the way to multidimensional probes utilizing longer excitation pulses and higher frequency resolution that will be even more sensitive to fluctuating order. In cuprates, future optical probes of fluctuations will be able to access fluctuations with full momentum selectivity and can be applied to more complex superconducting phases, for example to putative pair-density wave order in $La_{2-x}Ba_xCuO_4$ [32] or light-induced non-equilibrium phases in $YBa_2Cu_3O_{6+x}$ [33], toward elucidating their underlying microscopic mechanisms. More generally, our work further opens numerous experimental possibilities in quantum materials, and we envision that multidimensional terahertz spectroscopy will be applied to further outstanding problems concerning unconventional phase transitions [34, 35] as well as frustrated excitations inaccessible to linear spectroscopic methods [32, 36, 37].

# Methods

The sample used in our experiment was a single crystal of $La_{1.83}Sr_{0.17}CuO_4$, grown using the traveling-solvent floating-zone method. The crystal was cut and mechanically polished using diamond lapping film along an ac-oriented surface ~4 mm in size.

To perform multidimensional terahertz spectroscopy, two intense terahertz pulses were generated by optical rectification of 100 fs, 1300 nm pulses in two OH1 (2-{3-(4-hydroxystyryl)-5,5-dimethylcyclohex-2-enylidene}malononitrile) organic crystals. The two terahertz pulses are then focused in a non-collinear geometry onto the sample with a parabolic mirror of focal length 76.2 mm [38], resulting in peak electric fields of ~38 kV/cm and ~21 kV/cm at the sample surface (see also Supplemental Section S1) that together are comparable to the threshold electric field for reaching a nonlinear regime $E_{thresh} = (h\omega_p)/2ed \sim 40$ kV $cm^{-1}$ [39]. The emitted nonlinear electric field was then detected by conventional electro-optic sampling using 100 fs, 800 nm pulses in a ZnTe crystal. A differential chopping scheme, in which $E_A$ and $E_B$ were modulated at 500 Hz and 333 Hz respectively, was used to isolate the nonlinear electric field from the excitation fields.